# Intelligent Interference Management Based on On-Demand Service Connectivity for Femtocellular Networks

Kazi Nawshad Azam and Mostafa Zaman Chowdhury
Dept. of Electrical and Electronic Engineering, Khulna University of Engineering & Technology, Bangladesh
Email: nawshad_kuet@yahoo.com, mzceee@yahoo.com

*Abstract*—The femto-access-point (FAP), a low power small cellular base station provides better signal quality for the indoor users as to provide high data-rate communications with improved coverage, access network capacity and quality of service. Highly dense femtocellular deployments—the ultimate goal of the femtocellular technology—will require significant degree of self-organization in lieu of manual configuration as to mitigate unwanted handovers and interference effects. The deployment of femtocells with our proposed on-demand scheme in the multistoried building or home environment can solve the low level signal-to-noise plus interference ratio and throughput problems. In this paper, we propose new application of the femtocell technology with a new idea of femto-idle-mode and femto-active-mode system. The signal-to-noise plus interference ratio level and throughput are analyzed. The simulation results show that the proposed on-demand scheme in femtocell deployment significantly enhances the signal-to-noise plus interference ratio level and throughput in the multistoried building or home environment.

## I. Introduction

Now a day femtocellular network system has widened the horizon of new research areas. Femtocell is a low cost, more reliable, less power utilizing network deployment planning for future cellular system with a small coverage area [1]-[7]. The wireless engineering community has been searching for low-cost indoor coverage solutions since the beginning of mobile networks. Femtocellular network technology is one of such solutions. The femto-access-points (FAPs) enhance the service quality for the indoor mobile users. Some key advantages of femtocellular network technology are the improved coverage, reduced infrastructure and the capital costs, low power consumption, improved signal-to-noise plus interference ratio (*SNIR*) level at the mobile station (MS), and improved throughput. Femtocells operate in the spectrum licensed for cellular service providers thus, it can provide high performance. Also no need to use the dual mode terminal for this technology, whereas WLAN needs dual mode terminal and hence the key feature of the femtocell technology is that user do not require any new femtocell user equipment (FUE). One of the key advantages of the femtocellular technology is in the fact that it uses the same frequency bands as the macrocellular networks, thus avoiding the need to introduce new user equipment. However, the use of the same frequency spectrum can also cause substantial interference if no adequate interference management is incorporated into the network design, infrastructure, and the future extension plan.

Interference between two or more femtocells could be managed through on-demand scheme along with proper frequency allocation scheme, which would allow largest utilization of the valuable radio spectrum and the highest level of user's quality of experience (QoE). Specifically, appropriate interference management, implemented through our proposed scheme and suitable frequency allocation scheme, increases the system capacity, reduces the unwanted handovers and outage probability, and increases the frequency utilization. Although the interference in a femtocellular network cannot be fully eliminated, it is possible to reduce the interference to within a reasonable range by proper management.

In this paper we propose an effective femtocell to femtocell/macrocell and vice-versa handover scheme with a new idea of femto-idle-mode (FIM) and femto-active-mode (FAM) system. An FUE can be in any of three states namely idle state (only turn on), active state (turn on and attempting a call) and detached state (turn off). When FUE is in active state, only then FAP changes its mode from FIM to FAM. Therefore, the interference from several unnecessary FAPs can be reduced.

The rest of this paper is organized as follows. Section II shows the proposed on-demand service connectivity scheme. System model for the proposed scheme is shown in Section III. Section IV provides the *SNIR* level and throughput analyses. Finally, conclusions are drawn in the last section.

## II. Proposed On-Demand Scheme

Figure 1 shows three scenarios of a femtocell network deployment without on-demand scheme in integrated macrocellular network. In the first scenario, there is no femto user in FAP# 1 but FAP# 1 is active. There are two idle femto users in FAP# 2 but FAP# 2 is active in the second scenario. FAP# 3 is also active with a detached femto user in the third scenario of Figure 1(a). From these three scenarios it is obvious that in any condition FAPs are active. Besides this, in our proposed scheme a FAP is in FAM when there is a demand for it. As FAP activeness is depending on demand creation so that our proposed scheme is called on-demand.

Figure 2(a) shows that a FAP is not in FAM at every time because FAP activeness depends on demand creation. But demand creation depends on different situations. When a femto user requests or want to use the idle FAP or a handover request comes to an idle FAP then demand creates. After demand creation, a FAP become active.

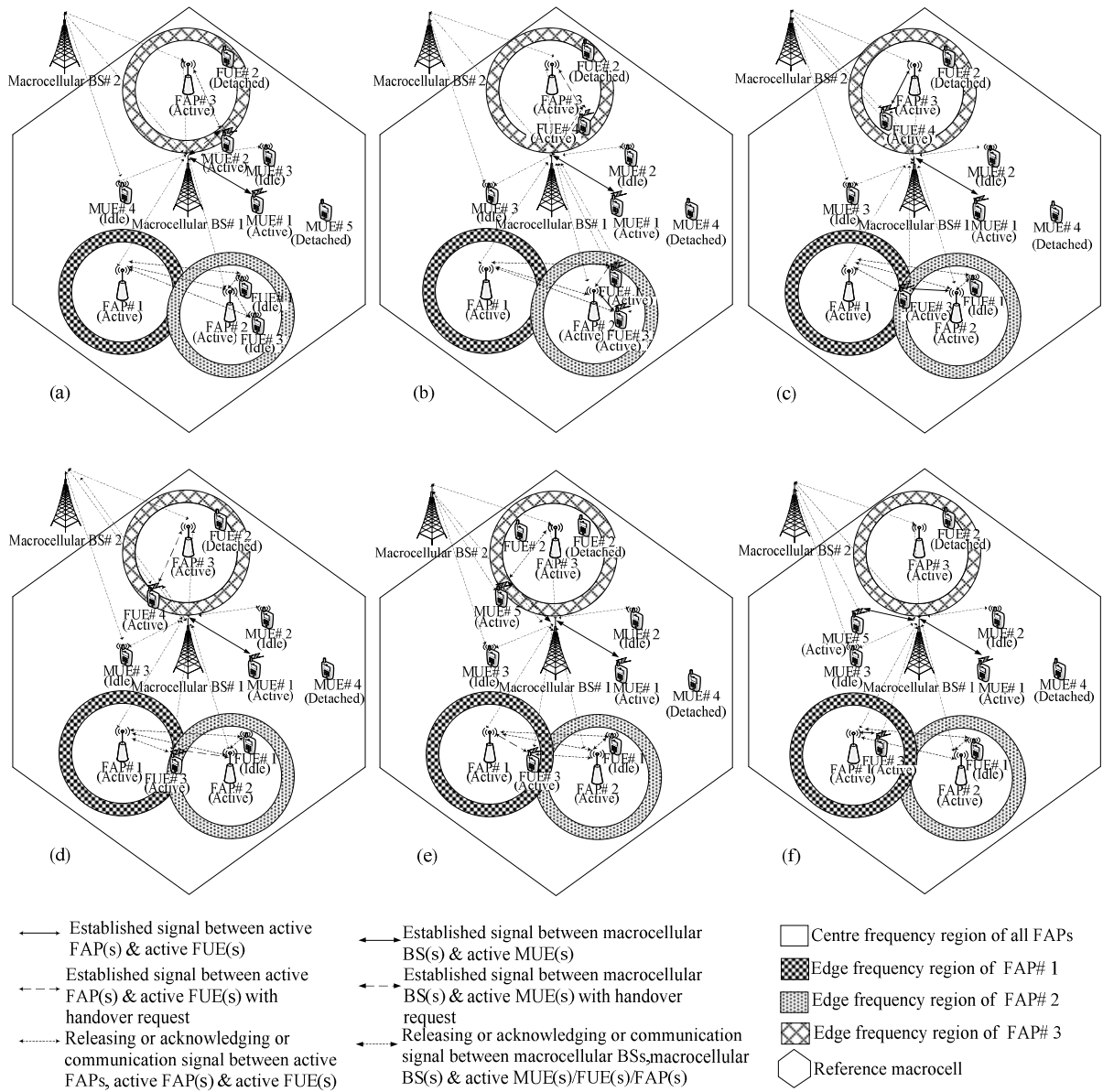

**Figure 1.** Example of successful femtocell-to-femtocell, femtocell-to-macrocell and macrocell-to-femtocell handovers of UEs in existing scheme (FAP with dynamic frequency-reuse scheme).

It is advantageous to keep a FAP in active and idle modes and one of the advantages is if we can keep lower number of active FAP then interference becomes low. Since FAP activeness depends on demand creation so that power consumption by a FAP can be reduced in our proposed scheme and hence it is called on-demand scheme. From Figure 2(a) it is clear that all FAPs are in idle mode because there is no demand. When an active femto/macro user wants to use any idle FAP or comes edge region of any idle FAP or sends request for a macrocell/femtocell-to-femtocell handover but not in direct connection with desired idle FAP then requested idle FAP becomes active. As FAP# 3 becomes active so it starts power transmission. But in idle mode, FAP# 3 transmits beacons only with regular interval instead of continuous power transmission. Again in our proposed scheme it is seen from Figure 2(a) that FAP# 1 has no femto user so no demand exists here and FAP# 1 is in FIM, FAP# 2 have two idle femto users so without any demand FAP# 2 is also in idle mode, FAP# 3 has a detached femto user so there is no demand and FAP# 3 remains in idle mode and in turns we can reduce interference by keeping active FAP in idle mode by creating no demand situation as explained earlier. If we observe Figure 2(a) minutely, it is found that MUE# 2 is in active state and connected with macrocellular base station (MBS)# 1 in reference macrocell and moving towards idle FAP# 3. While active MUE# 2 moves a certain distance to reach very near to edge region of idle FAP# 3, active MUE# 2 creates demand by creating a handover probability and idle FAP# 3 becomes active so that active FAP# 3 can take active MUE# 2 as active FUE# 4 which is shown in Figure 2(b). Figure 2(b) also shows that anyone idle femto user or both idle femto users of FAP #2 become active and makes idle FAP# 2 into active. Figure 2(c) depicts that one of the active femto users of active FAP# 2 of Figure 2(b) becomes idle and other active femto user is moving towards FAP# 1 but remains in active state so that FAP# 2 remains active. Here, FAP# 1 has no active femto user so that FAP# 1 remains in idle mode. But an active FUE# 3 moves towards edge of FAP# 1 and creates a handover probability so that FAP# 1 changes its mode from idle to active. When active FUE# 3 fully moved to FAP# 1 then FAP# 2 has only idle FUE# 1 and thus active FAP# 2 becomes idle. We can see from Figure 2(a) to Figure 2(f) that an idle FAP is in FAM if there is a demand for it. Since there are thousands of FAPs in a femtocell network deployment, interference can be reduced significantly if we can change several active FAPs into idle FAPs.

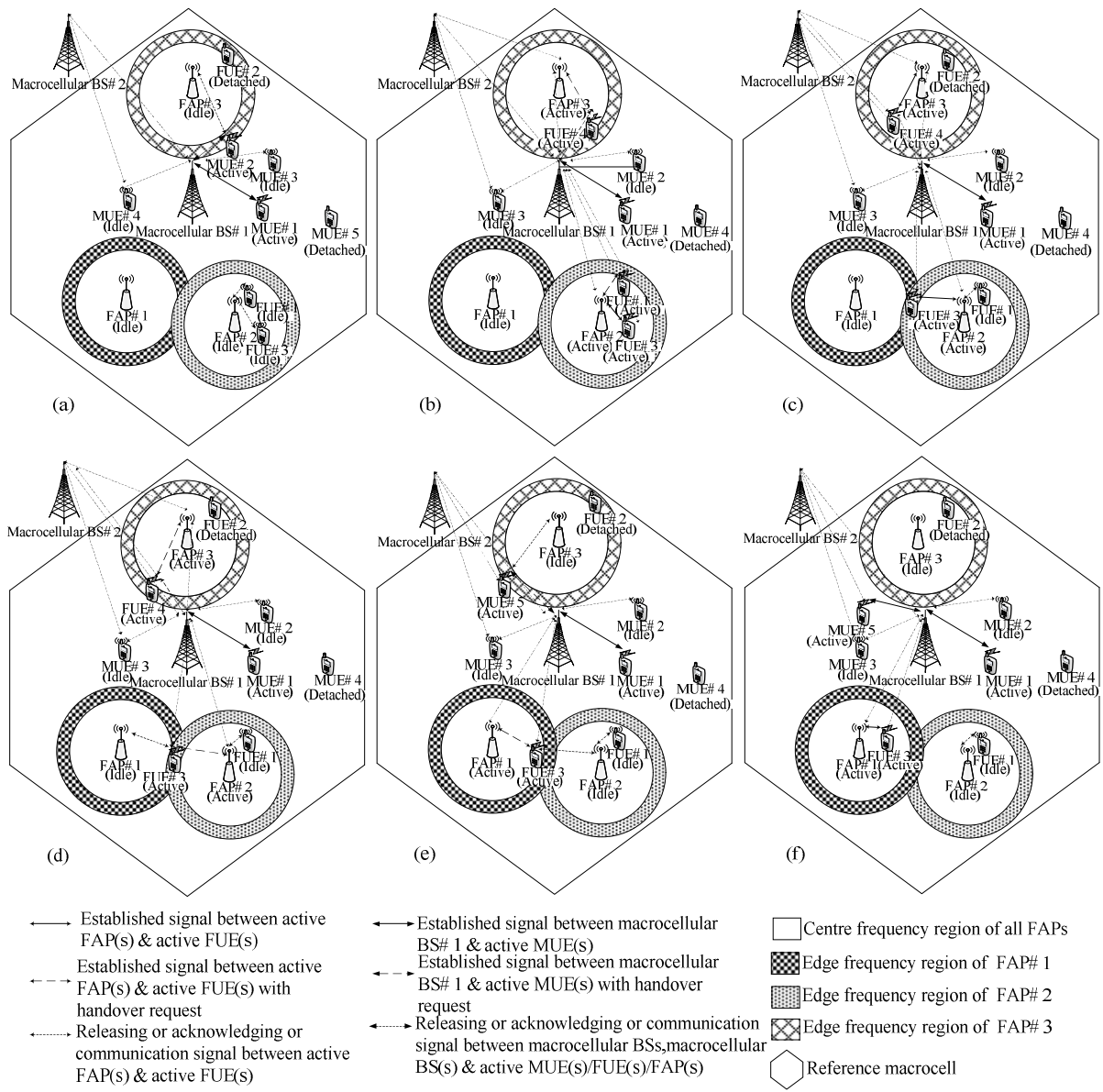

**Figure 2.** Example of successful femtocell-to-femtocell, femtocell-to-macrocell and macrocell-to-femtocell handovers of UEs in our proposed on-demand scheme with less interference effects (FAP with dynamic frequency-reuse scheme).

The basic flow diagram of the proposed scheme is given in Figure 3. Whenever a new FAP is ready to be deployed (or there is any active FAP) then at first the new FAP does its system parameter selection job as to recognize its neighbor environment. Then the FAP selects its type as it is in standalone or dense femtocell environment from neighbor environment. Sizing the system/network area and own coverage area is done by the FAP from the necessary measurements. Then femtocell-to-femtocell (in dense femtocellular system only) or femtocell-to-macrocell (in standalone/dense femtocellular system), interference analysis is done with the help of FGW and neighbor environment. When an active femto user exists in the new FAP coverage area then the link quality between femto user and FAP, between FAP and FGW etc. are checked in order to response to a handover request of serving active FAP through FGW for a successful handover between active FAPs or active FAP and macrocell. Now all active neighbor FAP list (in dense femtocellular system only), all active neighbor FAP signal strength and all active femtocell user (under serving active FAP) distance monitoring, sharing and updating are done (under same FGW) between the FGW and active FAP and between the FGW and all neighbor active FAP in order to avoid unwanted handovers and high interference effects.

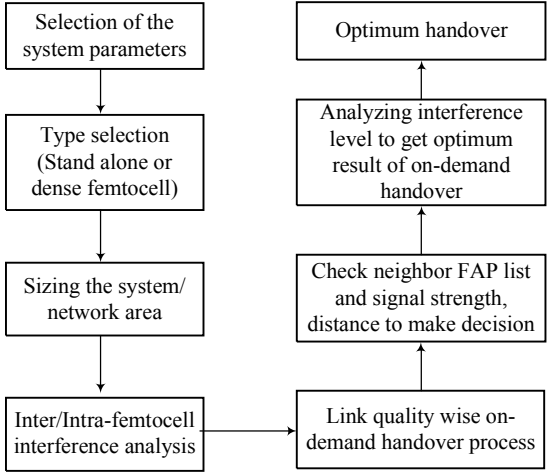

**Figure 3.** Flow diagram of on-demand handover mechanism.

When active femtocell user satisfies all conditions of the femtocellular system relating to successful handover process under serving active FAP then active FAP is ready for initializing a handover with the help of FGW else handover process aborts. Finally, optimum handover is done with lower interference effect. After a successful handover, the whole process repeats for other active femtocell user or active FAP goes to idle mode if no active femtocell user exists therein.

## III. System Model

There are various interference sources present in the femtocell network; in particular, among femtocells, and between femtocells and macrocells [3]. In addition, the noise is an element of the wireless environment. These interferences and the noise affect the capacity of the wireless link. In this section, we provide the *SNIR* level and throughput analyses for the femtocellular network deployment.

The path loss equation used for reference femtocell is

$$L_r = 20\log_{10}(f_{ue}) + N\log_{10}(d_r) - 28 \quad (1)$$

where $L_r$ is path loss exponent for reference FAP, $f_{ue}$ is centre frequency of the femtocell, $d_r$ is distance between the reference FAP and the active FUE, $N$ is total power of the received noise.

The path loss equation used for neighbor femtocell is

$$L_i = 20\log_{10}(f_{ue}) + N\log_{10}(d_i) + 4(n_i)^2 - 28 \quad (2)$$

where $L_i$ is path loss exponent for *i*-th active neighbor FAP, $d_i$ is the distance between *i*-th neighbour FAP and the active FUE, $n_i$ is number of walls between the FUE and the *i*-th neighbour FAP.

The equation of received power for reference femtocell is

$$P_R = P_0 10^{(\frac{-L_r}{10})} \quad (3)$$

where $P_R$ is power received by reference FAP, $P_0$ is power of the signal from the reference FAP.

The equation of *SNIR* used for reference femtocell is

$$SNIR_r = \frac{P_R}{(I_{fr} + I_m + N)} \quad (4)$$

where $I_{fr}$ is total power of the received interference from all interfering idle FAPs, $I_m$ is total power of the interference from all of the interfering macrocells, $SNIR_r$ is signal-to-noise plus interference ratio level for reference FAP.

The equation of *SNIR* used for neighbor femtocell is

$$SNIR_i = \frac{P_R}{(I_{fi} + I_m + N)} \quad (5)$$

where $I_{fi}$ is total power of the received interference from *i*-th interfering active FAP, $SNIR_i$ is signal-to-noise plus interference ratio level for *i*-th active neighbour FAP.

Assuming that the spectrums of the transmitted signals are spread, we can approximate the interference as AWGN. The Shannon equation for capacity is known as

$$C = W\log_2(1 + SNIR) \quad (6)$$

where $C$ is the capacity used in Shannon equation, $W$ is the bandwidth used in Shannon equation, *SNIR* is signal-to-noise plus interference ratio level used in Shannon equation.

## IV. Simulation Results

In this section we evaluate the *SNIR* level and throughput of the proposed scheme. Table 1 summarizes the values of the parameters that we used in our analysis. We assume that the femtocells in the multistoried building or home environment and the overlaid macrocell are deployed through dynamic frequency-reuse scheme.

**Table 1.** Summary of the parameter values used in analysis.

| Parameter name | Value |
|---|---|
| Distance between the FAP centre and outer boundary region | 10 m |
| Distance between the FAP and active FUE | 5 m |
| Distance between two FAP centers | 20 m |
| Carrier frequency | 1800 MHz |
| Transmit signal power by the macrocellular BS | 1.5 kW |
| Transmitted signal power by a FAP | 15 mW |
| Height of a macrocellular BS | 100 m |
| Height of a FAP | 2 m |
| Threshold value of *SNIR* at FAP inner boundary region for active FUE ($\gamma_i$) | 12.55 dB |
| Threshold value of *SNIR* at FAP outer boundary region for active FUE ($\gamma_o$) | 8.21 dB |
| Noise power ($N$) | 6.9882 x $10^{-7}$ mW |
| Number of walls between the active FUE and the *i*-th neighbor FAP ($n_i$) | 1 (for 1st tier), 2 (for 2nd tier) |
| Number of neighbor FAPs | 30 |

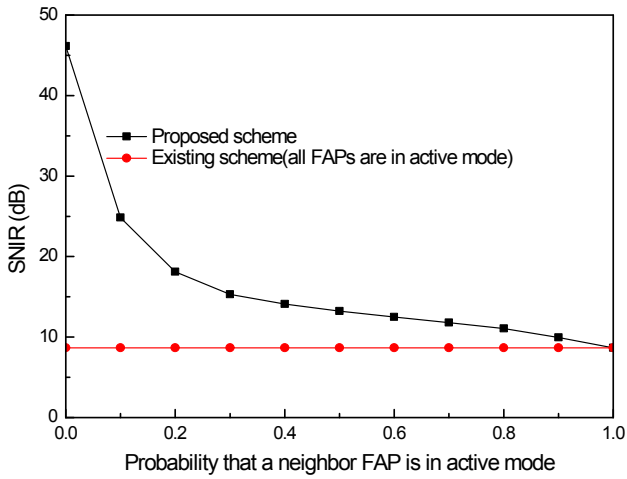

**Figure 4.** *SNIR* analysis with probability that a FAP is in active mode.

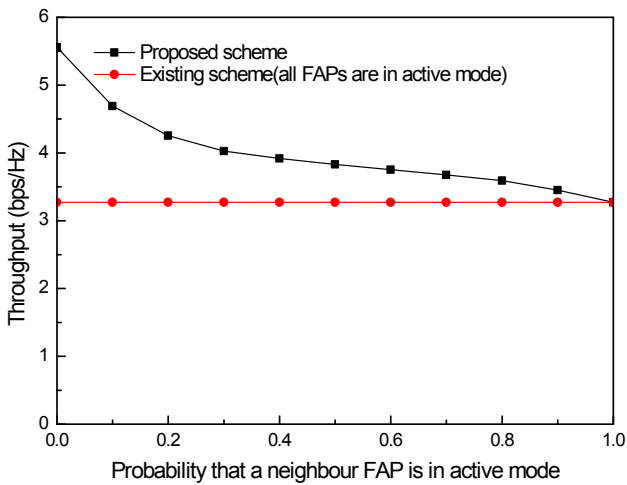

**Figure 5.** Throughput analysis with probability that a FAP is in active mode.

Figure 4 shows *SNIR* analysis with probability that a FAP is in active mode. Probability is zero means no neighbor FAP is in active mode in our proposed scheme. For instance, probability factor 0.4 indicates total number of active FAPs is 12 out of 30 neighbor FAPs. When all FAPs are in active mode i.e. existing scheme then probability factor is always unity. From the Figure 4, we can see that when probability factor is zero then *SNIR* is at maximum level for our proposed scheme and for existing scheme its value is always constant. At the unity probability factor, the *SNIR* for both cases are equal. The performance curve of our proposed scheme is always higher than other performance curve but it intersects only at unity probability factor point. More *SNIR* means better the signal strength for serving any active FUE and it varies in our proposed model in accordance with probability of neighbour active FAPs. So, lower interference effect can be achieved.

Figure 5 shows throughput analysis with probability that a neighbour FAP is in active mode. Performance curve of our proposed scheme is higher than the performance curve where all FAPs are always active. Throughput of our proposed scheme varies in accordance with number of active neighbor FAPs but for existing scheme it is constant. More throughput means higher capacity per head user and hence it varies in our proposed scheme in accordance with probability of active FAPs. So, better QoS can be ensured.

Figure 6 shows *SNIR* analysis with number of active FAPs. In our proposed scheme maximum and minimum *SNIR* occurs at zero and fifteen (randomly chosen) active neighbor FAP point, respectively depicts the wide range of throughput variation. But the performance curve with existing scheme is a straight line with a constant throughput value. The performance curve of our proposed scheme is always above the other curve. The higher value of *SNIR* in our scheme provides less interference effect in serving any active FUE by reference active FAP.

Figure 7 shows throughput analysis with number of active FAPs. The throughput performance curve follows the *SNIR* performance curve shown in Figure 6. Hence, in our proposed scheme maximum throughput per head user gives femtocell user connectivity with higher capacity. So, better QoS can be ensured with proposed on-demand scheme.

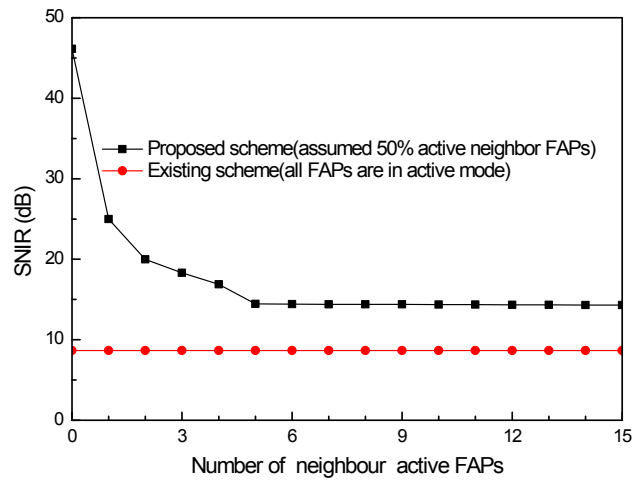

**Figure 6.** *SNIR* analysis with number of active FAPs.

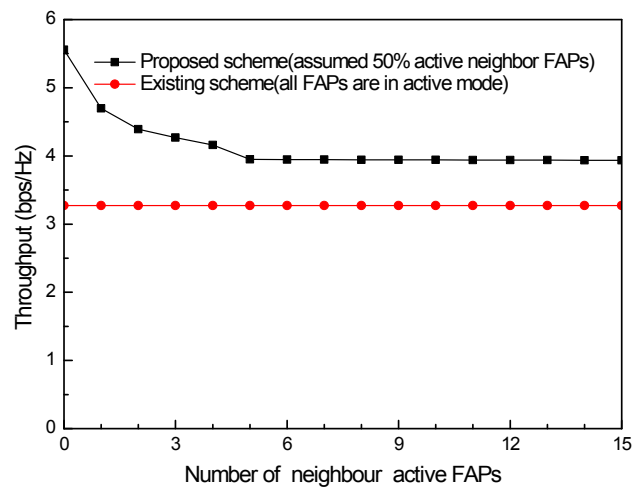

**Figure 7.** Throughput analysis with number of active FAPs.

All four performance curves in Figures 4 - 7 show that the radio resource used is variable in accordance with number of active FAPs. So, intelligent interference management system is ensured efficiently because when all FAPs are in FIM then very low power is consumed by all FAPs.

## V. Conclusions

Femtocellular network system is a small and low cost alternative for meeting the demand of increasing data traffics in existing cellular networks. With the proposed on-demand service connectivity scheme – interference effects can be managed intelligently than existing management techniques because here we proposed a new approach, where a FAP has femto-idle-mode and femto-active- mode operations. While FAPs are in idle mode, all the FAPs uses very low power and in the active mode case we considered here probability factor and compare our proposed scheme performances with existing scheme performances (always 100% neighbor FAPs are active) for *SNIR* and throughput calculation. For all cases our proposed scheme performance shows better performance and hence ensures higher QoS, less power consumption due to two operational modes of FAP, which is an intelligent way to less interference effect and lower unwanted handover by on-demand handover scheme. From the analysis, it is obvious that power consumption of FAPs is reduced significantly with our proposed on-demand handover scheme. Therefore, our proposed scheme is a promising approach for the dense deployment of femtocellular networks.